\documentclass[10pt,letterpaper]{article}
\usepackage{opex3}
\usepackage{cite} 

\usepackage[ocgcolorlinks]{hyperref}
\usepackage{amsmath}
\usepackage{amssymb}

\newcommand{\Fi}{\mathcal{F}}
\newcommand{\Lcav}{L}
\newcommand{\mmeff}{\epsilon} 

\begin{document}

\title{Millimeter-long Fiber Fabry-Perot cavities}

\author{Konstantin Ott$^{1,2}$, Sebastien Garcia$^{1}$, Ralf Kohlhaas$^{2}$, Klemens Sch\"uppert $^{3}$, Peter Rosenbusch$^{2}$, Romain Long$^{1}$ and Jakob Reichel$^{1,*}$}

\address{$^1$Laboratoire Kastler Brossel, ENS-PSL/CNRS/UPMC-Sorbonne
  Universit\'es/CdF,\\24 rue Lhomond, 75005 Paris, France\\
$^2$ LNE-SYRTE, Observatoire de Paris, CNRS/UPMC-Sorbonne
  Universit\'es,\\61 av. de l'Observatoire, 75014 Paris, France\\
$^3$ Institut f\"ur Experimentalphysik, Universit\"at Innsbruck,\\Technikerstra{\ss}e 25, 6020 Innsbruck, Austria}

\email{$^*$jakob.reichel@ens.fr}


\begin{abstract}
  We demonstrate fiber Fabry-Perot (FFP) cavities with concave mirrors
  that can be operated at cavity lengths as large as 1.5\,mm without
  significant deterioration of the finesse. This is achieved by using
  a laser dot machining technique to shape spherical mirrors with
  ultralow roughness and employing single-mode fibers with large mode
  area for good mode matching to the cavity. Additionally, in contrast
  to previous FFPs, these cavities can be used over an octave-spanning
  frequency range with adequate coatings. We also show directly that
  shape deviations caused by the fiber's index profile lead to a
  finesse decrease as observed in earlier attempts to build long FFP
  cavities, and show a way to overcome this problem.
\end{abstract}


\bibliographystyle{osajnlJR}
\bibliography{LongFFPs,LongFFPsLocal}

\section{Introduction}

Fiber Fabry-Perot (FFP) microcavities with CO$_2$ laser-machined concave
mirrors \cite{Colombe07,Hunger10b} are being used in a fast-growing
number of applications, ranging from cavity quantum electrodynamics
with atomic, molecular and solid-state emitters
\cite{Colombe07,Toninelli10,Steiner13,Albrecht13,Miguel13,Besga15} and
optomechanical systems \cite{Flowers12} to Raman spectrometers for
atmospheric gases \cite{Petrak14} and new scanning microscopy
techniques \cite{Mader15}. This range of applications could be further
increased, and a gap in micro-optical technology could be filled, by
increasing the optical length $\Lcav$ of these cavities to the
millimeter range while maintaining their crucial advantages such as
built-in fiber coupling, small mode waist, high finesse and high
passive stability. Development towards this goal has been initiated by
the ion trap community \cite{Brandstaetter13,Takahashi14}, where FFP
cavities are now arguably the most promising candidate for realizing a
high-fidelity light-matter quantum interface. In this application,
increased cavity length will mitigate or even remove the perturbation
of the trapping potential induced by the dielectric mirrors. With
neutral atoms, increased mirror distance will allow elongated atomic
ensembles to be placed in the cavity, as required for improving
compact, trapped-atom atomic clocks \cite{Szmuk15} by cavity squeezing
\cite{Leroux10b,Hosten16}. Finally, because of the reduced free
spectral range, larger $\Lcav$ will also make it possible to use
laser-machined concave mirrors in telecom applications of FFP
cavities, where planar mirrors \cite{Miller90} are currently being
used in commercial solutions \cite{MicronOptics}.

Two factors have limited $\Lcav$ in earlier FFP
implementations. First, the nonspherical profile generated by a single
CO$_2$ laser pulse causes additional losses which are negligible for
short cavities, but become important long before $\Lcav$ reaches the
stability limit
\cite{Hunger10b,Brandstaetter13,Takahashi14,Benedikter15}. The
Gaussian rather than spherical shape \cite{Hunger10b} limits the
effective mirror diameter, even when a very large CO$_2$ beam diameter
is used \cite{Uphoff15}. Shape deviations due to the doping of the
fiber core \cite{Takahashi14} and to the restricted heat flow in the
fiber \cite{Hunger12} also contribute to this limitation.  Second,
because FFPs have no mode-matching optics, the mode field diameter of
the fiber determines the coupling efficiency between the fiber and
cavity mode. With typical mirror curvatures, the standard single-mode
(SM) fibers used in earlier FFP implementations provide near-perfect
mode matching to short cavities, and are used with great success in
cavity quantum electrodynamics experiments. Resonant cavity
transmission $T_c$ rapidly drops, however, when $\Lcav$ is increased
beyond a few hundred $\,\mu$m \cite{Hunger10b,Brandstaetter13}.

Concerning the mirror profile, a natural approach is to use multiple
CO$_2$ laser pulses on the same fiber, instead of a single pulse. A
first application of this approach was demonstrated recently in
\cite{Takahashi14}, where
the fiber was rotated about its axis between pulses. The main purpose
of this rotation was to reduce mirror ellipticity, but the technique
also resulted in larger mirror profiles, enabling values of $\Lcav$ up to
$400\,\mu$m before the finesse dropped by 50\%, for cavities with one
SM fiber. However, the mirror profile still remained Gaussian and the
problem of reduced transmission remained unsolved.

Here, we overcome both limitations. To address the mirror shape
limitation, we apply a newly developed CO$_2$ dot milling technique,
where a large number of weak individual pulses sequentially address an
optimized pattern of target points on the substrate surface.  This
method gives access to a wide range of shapes with extremely precise
control over the surface profile, while maintaining the excellent
surface roughness that is characteristic of CO$_2$ machining. Here we
show that it enables fabrication of large, spherical structures, as
required for long cavities, with a shape deviation of less than
$\lambda/20$. To address the mode-matching issue, we use
large-mode-area photonic-crystal (PC) fibers, also known as
``endlessly single-mode'' fibers \cite{Birks97}.  We find that CO$_2$
machining produces excellent results on these fibers when the holes
are collapsed adequately.  To measure the performance of the improved
mirrors, we have built cavities and performed finesse and transmission
measurements. $\Lcav>1.6\,$mm is reached before the finesse drops by
50\%.  We present measurements of finesse as a function of $\Lcav$ for
different cavities and compare them to simulations that use the
measured mirror profiles, finding good agreement. These simulations
also clarify the impact of doping-related shape variations. We also
measure cavity transmisson over the full length range and directly
compare the results of a PC-multimode (MM) fiber cavity to a SM-MM
cavity with similar mirror parameters. We find that the PC fiber
improves the transmission by an order of magnitude for cavity lengths
beyond 1\,mm.

\section{The role of the cavity mode radius on the fiber mirrors}

The mode radius of the cavity mode on the mirrors plays an
important role for both the losses and the mode matching. Limiting
ourselves to symmetric cavities for simplicity, the mode radius on the
cavity mirrors is $w_m=w_c\sqrt{1+\left(L/(2z_R)\right)^2}$
where $w_c=\sqrt{\lambda/(2\pi)}(\Lcav(2R-\Lcav))^{1/4}$ is the waist radius
of the cavity mode and $z_R=\pi w_c^2/\lambda$ its Rayleigh range, $R$
is the radius of curvature (ROC) of the mirrors, $\Lcav$ the cavity
length, and $\lambda$ the wavelength. A small $w_m$ has the advantage of
minimizing the clipping losses due to the finite mirror diameter (see
Sec. \ref{sec:LongFFPs} below). Additionally, for the long cavities
that we are targeting here, $w_m$ tends to be larger than the mode
field radius of the input fiber, $w_f$. This reduces the power
coupling efficiency $\mmeff$ from the fiber to the cavity, and
consequently, the overall (fiber-to-fiber) resonant cavity
transmission $T_c$. Here as well, reducing $w_m$ is beneficial. The
smallest $w_m$ that can be reached for a given $\Lcav$ is realized for
$R=\Lcav$, i.e., a confocal cavity. In that case,
\begin{equation}
  w_{m,\min}=\sqrt{\frac{\lambda}{\pi}\Lcav}\ \ and\ \  w_{c}=\sqrt{\frac{\lambda}{2\pi}\Lcav}.\
\end{equation}
Taking $\lambda=780\,$nm
and $\Lcav=1.2\,$mm for example, we find $w_{m,\min}=17.3\,\mu$m and $w_{c}=12.2\,\mu$m. This
is still much larger than the mode field radius of a typical SM
fiber, so that a better solution must be found, as discussed further
below.

The fiber-to-cavity power coupling efficiency can be approximated by the mode
overlap of two Gaussian modes \cite{Joyce84},
\begin{equation}
  \mmeff \approx \frac{4}{\left(\frac{w_f}{w_c}+\frac{w_c}{w_f}\right)^{2}+\left(\frac{\lambda}{\pi w_f w_c}\right)^2 s^2}\,,
\label{eq:powerTransmissivity}
\end{equation}
$w_f$ being the waist radius of the mode entering the cavity from the
fiber, $w_c$ that of the cavity mode as before, and $s$ the distance
between the two waist positions.  The lensing effect of the concave
mirror structure, as well as the additional phase mismatch due to the
wavefront curvature of the cavity mode can be neglected for the long
FFPs considered here. (A treatment including these effects can be
found in the appendix of \cite{Hunger10b}.) For a symmetric cavity,
$s=\Lcav/2$. We find it convenient to introduce the dimensionless
factor
\begin{equation}
\alpha\equiv \frac{\Lcav}{2z_R}=\frac{1}{\sqrt{2\frac{R}{L}-1}}\,,
\end{equation}
such that $\alpha=1$ for a confocal cavity, $\alpha\to 0$ for a short
cavity, and $\alpha\to\infty$ when approaching the concentric limit
$\Lcav=2 R$.  The power coupling efficiency can then be written
\begin{equation}
  \mmeff = \frac{4}{\left(\frac{w_f}{w_c}+\frac{w_c}{w_f}\right)^{2}+\left( \frac{w_c}{w_f} \right)^2\alpha^2}\,,
  \label{eq:powerTransmissivityAlpha}
\end{equation}
which depends only on the ratio $w_f/w_c$ and $\alpha$.

\begin{figure}[htbp]
\centering
\includegraphics[width=13cm]{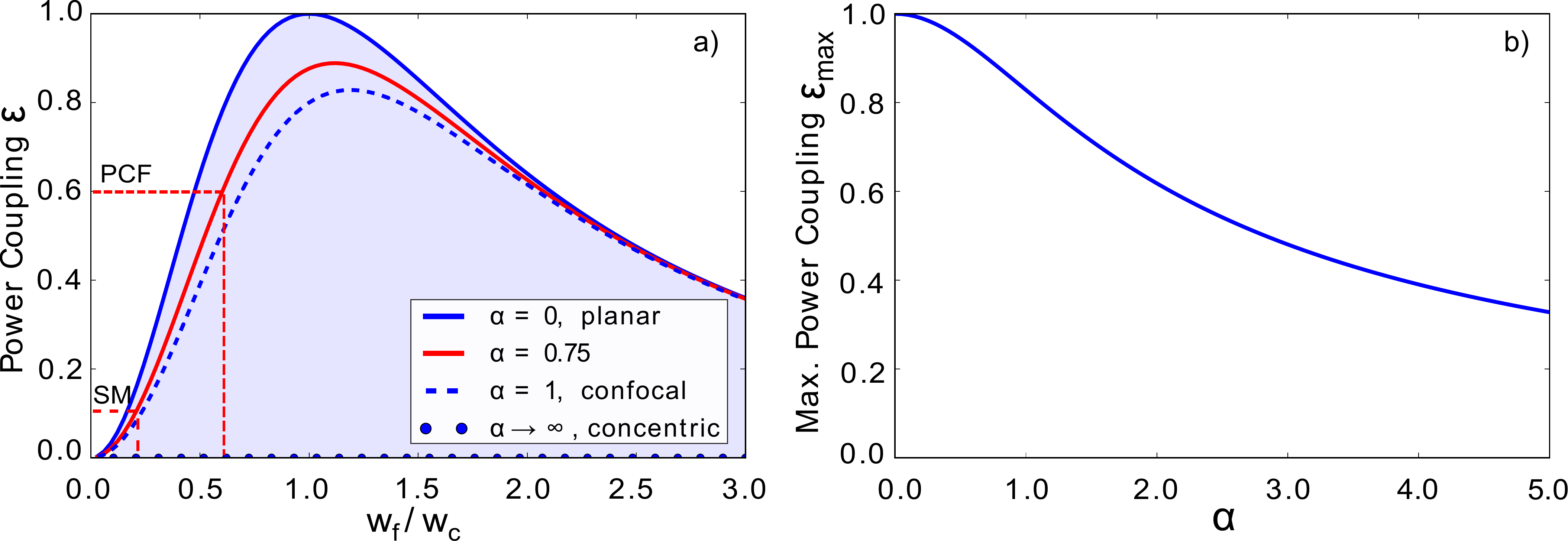}
\caption{a) Power coupling between the incoupling fiber mode and the
  mode of a symmetric cavity for different values of $\alpha$ (the
  cavity length normalized to the Rayleigh range of the cavity mode).
  The red dashed lines indicate the increased power coupling expected
  from the use of a PC fiber ($w_f=8.2\,\mu$m) with respect to a
  standard SM fiber ({$w_f=3\,\mu$m}) for an example cavity with
  $w_0=14.1\,\mu$m and $L=1.2\,$mm. The blue shaded area shows the
  stability region of the symmetric cavity. b) Maximum achievable
  power coupling for a given $\alpha$.}
\label{fig:SMvsPCF}
\end{figure}

Fig.~\ref{fig:SMvsPCF}~a) shows this coupling efficiency for different
$\alpha$. The optimum coupling $\epsilon_{max}$ for a given $\alpha$
is reached for $w_f/w_c=\sqrt[4]{1+\alpha^2}$. For configurations with
$0\le \alpha\lesssim 1$, this corresponds to $1\le w_f/w_c \lesssim 1.2$:
for cavity lengths up to the confocal length, the optimum fiber mode
radius is always close to the waist of the cavity mode. 
A coupling efficiency rigorously equal to 1 can only be reached for
$\alpha=0$, but high coupling efficiencies $\mmeff>0.8$ are possible
for a wide range of $\alpha$ values, as shown in
Fig.~\ref{fig:SMvsPCF}~b). Whether these efficiencies can actually be realized
experimentally depends on the availability of a suitable fiber. For
long cavities, we have already seen that $w_m$ tends to be larger than
typical fiber modes, and can also cause losses due to finite mirror
diameter. Therefore, a good working point for these cavities is close
to the confocal configuration $\alpha=1$, where $w_m$ is smallest.
Note that further optimization is possible in situations where
asymmetric cavities can be employed, notably for plano-concave
configurations with asymmetric reflectivity.

\section{Photonic-Crystal Fibers for FFP Cavities}	
\label{sec:PCF}

The discussion above shows that mode matching for long fiber cavities
can be substantially improved by using a fiber with larger mode field
diameter. It is known that large-core multimode fibers improve the
transmission (see \cite{Takahashi14} for a direct
comparison). However, the need for a well-defined, stable coupling
usually makes them a bad choice on the input side of the
cavity\footnote{In principle, this problem can be overcome by the use
  of time-reversal techniques to achieve the desired mode at the
  output of the multimode fiber \cite{Papadopoulos13}. However, a
  suitable scheme to obtain an error signal remains to be found.}. A
promising alternative, which we explore here, is to use single-mode
fibers with large mode area. Such fibers are available based on
different technologies. Here we use photonic-crystal (PC) fibers.

Let us evaluate the power coupling that can be expected for
$L=1.2\,$mm, which is the targeted cavity length in our application.
In practice, at the confocal point itself, stability is compromised by
the fact that the mirror ROCs have some tolerance from sample to
sample (see sec.~\ref{sec:Optimizing} below), so we choose to work at
$\alpha\sim 0.75$. The required ROC to realize a given combination
$(\alpha,\Lcav)$ is $R= \Lcav/2\left( 1/\alpha^2+1\right)$.  For
$\Lcav=1.2\,$mm and $\alpha=0.75$, the required ROC is $R=1.67\,$mm.
The calculated cavity mode waist for these parameters is
$w_c=14.1\,\mu$m at 780\,nm.  The red dashed lines in
Fig.~\ref{fig:SMvsPCF} show the expected power coupling efficiency of
this mode to a standard SM fiber with $w_f=3\mu$m,
$\mmeff\approx11\%$, and to the NKT LMA20 photonic crystal fiber which
we have used in our experiments, which has a specified near-field mode
field radius of {$w_f=(8.2 \pm0.8)\mu$m} at 780\,nm (see
Tab.~\ref{tbl:fibers} below for more information). This $w_f$ value
leads to $\mmeff\approx58\%$, nearly a factor 6 higher than with the
SM fiber. Note however that this idealized calculation neglects
effects such as the non-Gaussian shape of the PC fiber mode, so its
result should be considered as an upper limit.


In addition to the improved mode coupling, these PC fibers are
``endlessly single mode'', allowing stable single-mode guiding over a
large wavelength range which can span more than an octave. Here, we
will use them with a dual-wavelength coating optimized for 780\,nm and
1560\,nm.

\section{Machining large spherical structures by CO$_2$ dot milling}
\label{sec:MultipleShots}

\subsection{Limiting factors in single-pulse machining}
The profile of the depression generated by a single CO$_2$ laser pulse
is Gaussian rather than spherical \cite{Hunger10b}, so that the
effective mirror diameter is significantly smaller than the diameter
of the depression. Additionally, due to boundary effects in heat
transport during the laser machining, it is difficult to machine
depressions that extend over the full fiber surface \cite{Hunger12},
even with a very large CO$_2$ beam diameter \cite{Uphoff15}.
The small effective mirror size causes clipping loss when the transverse
mode size on the mirror becomes comparable to the effective mirror
diameter. This typically happens 
long before $\Lcav$ reaches to the stability limit
\cite{Hunger10b,Brandstaetter13,Takahashi14,Benedikter15}. Finally, it is
known that the milling profile of a given laser pulse is affected by
the doping profile of the fiber \cite{Takahashi14}. The effect is most
pronounced for SM fibers, where it leads to a circular ridge at the
core-cladding interface. All three problems can be addressed by
replacing the single CO$_2$ pulse by a series of pulses with different
target positions.

\subsection{Dot milling setup}
\label{sec:CO2machining}

\begin{figure}
\centering\includegraphics[width=13cm]{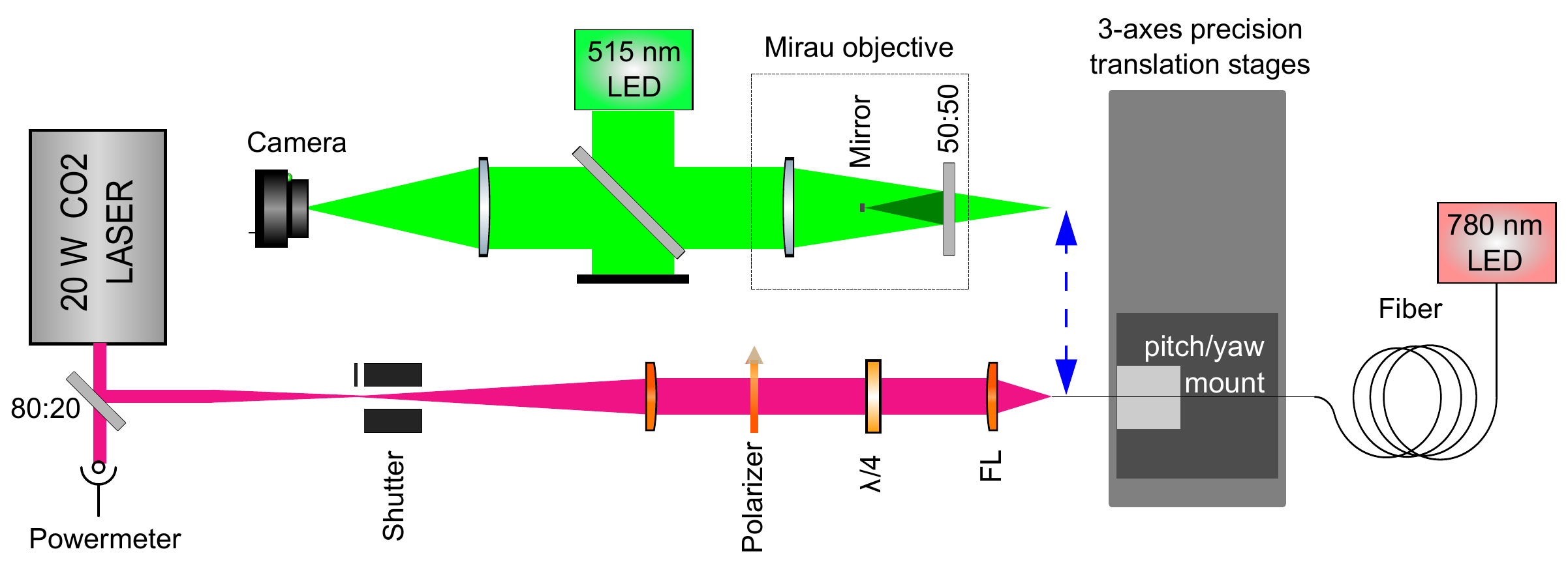}
\caption{(a) CO$_2$ dot milling setup: Precision translation stages
  are used to control the CO$_2$ beam position on the fiber and to
  translate the fiber to the optical profiler position for surface
  characterization. Light coupled into the fiber core allows a precise
  centering of the mirror profile. The CO$_2$ focusing lens (FL) is a
  $f=25\,$mm asphere illuminated with a beam diameter of 7\,mm.}
\label{fig:setup}
\end{figure}

Our machining setup, which will be described in more detail elsewhere
\cite{Garcia15}, is designed to enable surface machining with
multiple, precisely positioned CO$_2$ pulses (``shots'') and analysis
of the results on the fly. It has three main components: the CO$_2$
laser with external pulsing and focusing optics, a high-precision,
three-axis translation stage, and an in-situ optical profilometer
(Fig.~\ref{fig:setup}). The complete setup is enclosed in a
temperature-stabilized laminar flow box. 
A motorized tilt stage (pitch and yaw) was mounted on the translation stages 
to align the fiber cleave surface perpendicular to the CO$_2$ beam axis before machining.
The CO$_2$ part is similar to our earlier setups
\cite{Hunger10b,Hunger12}. However, the CO$_2$ and profilometry beam
paths are kept separated, with the translation stages ensuring
reproducible travel between the two locations. The
stages
have optical encoders for repeatble absolute positioning. The step
size of the transverse stages is $50\,$nm. The optical profilometer is
based on a Mirau objective
with a nominal transverse resolution of $0.8\,\mu$m. To achieve best
height ($z$) resolution, we use phase-shifting profilometry
\cite{Schmit95}. The measured noise for a reference surface is
$1.4\,$nm rms for a single profile, which can be further reduced to
the sub-nm level by averaging several profiles. To localize the fiber
core of SM fibers, weak light is coupled into the fiber, so that the
core appears as a bright spot on the microscope image. The whole
system is software-controlled and allows for automated centering.

With this improved setup, the measured average centering error of a
CO$_2$ shot with respect to the fiber core is $0.9\,\mu$m, which is a
significant improvement over the earlier setup, and just sufficient to
achieve reproducible mode-matching between the fiber and cavity
modes. The relative positions of the multiple shots making up the dot
milling pattern have much better accuracy, as they do not require
large displacements.

\subsection{Fiber preparation}
\label{sec:fiberPrep}

\begin{figure}[htbp]
\centering\includegraphics[width=12cm]{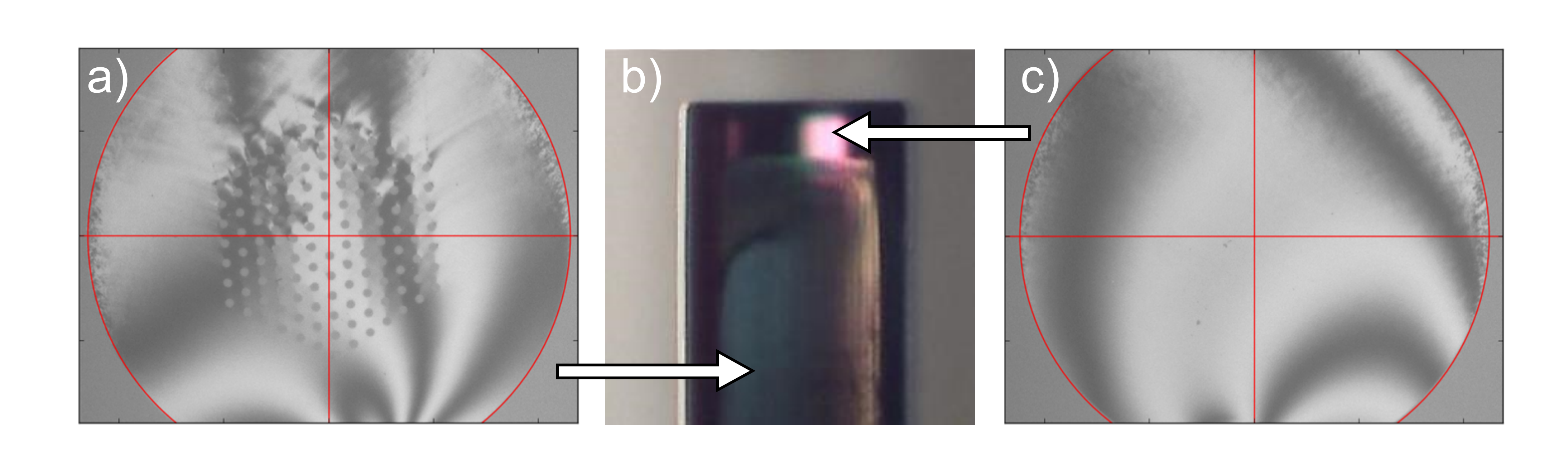}
\caption{(a) Profilometer image of the surface of a PC fiber (NKT LMA
  20) cleaved in the non-collapsed region, indicated by the arrow. (b)
  Side view of a PC fiber, showing the transition between the
  non-collapsed and collapsed regions.  (c) Profilometer image of the
  surface for a cleave in the collapsed region. (The red crosshairs in
  (a) and (c) mark the center of the fiber.)}
\label{fig:PCFcleaving}
\end{figure}

We have machined three different fiber types, presented in Table
\ref{tbl:fibers}.  For good reproducibility of the machining results,
it is important to control the cleave angle to tight tolerances. We
have used a pneumatic cleaver with a specified typical cleave angle of
less than 0.3 degrees \cite{Autocleaver}. Our measured mean
cleave angle deviation was less than 0.4 degrees throughout, and below
0.2 degrees for the SM fibers. To achieve these results, the cleaver
has to be calibrated carefully and the copper-coated fibers have to be
straightend by hand before cleaving. For the PC fiber, simple cleaving
exposes the holes of the photonic crystal. We found that direct
machining on that structure was possible for profiles with $R\lesssim
300\,\mu$m. For the much shallower profiles of our targeted
$R>1000\,\mu$m mirrors, this simple approach turned out to be
problematic.  Therefore, we have collapsed the holes before cleaving
using the arc of a fusion splicer as described in
Ref.~\cite{Colombe14}. The cleave was positioned at a distance
$d_c\sim 50-100\,\mu$m from the onset of the collapsed region as shown
in Fig.~\ref{fig:PCFcleaving}. Excellent results were achieved with
the fibers prepared in that way. For shorter $d_c$, residues of the
six-fold symmetry of the PC hole pattern remained visible in the
machined structure.

\begin{table}[h]
\centering
\begin{tabular}{|l|l|l|r|r|l|}
\hline
   \textbf{Type} &\textbf{Supplier} &\textbf{Ref.} & \textbf{MFD} &\textbf{Wavelength range} &\textbf{Coating}\\
&&&\multicolumn{1}{|c|}{$\mu$m}&\multicolumn{1}{|c|}{nm}&\\
\hline
	MM&IVG Fiber&Cu200/220 IR&200$\pm3$&700--1700&copper\\
	\hline
	SM&IVG Fiber&Cu800-200&     6$\pm0.5$&800--1000&copper\\
	\hline
	PC&NKT Photonics&LMA 20&16.4$\pm1.5$&700--1700&acrylate\\
\hline
\end{tabular}
\caption{Fiber types used in the experiments. The mode field
  diameter (MFD) indicated here is the nominal value specified by the
  manufacturer. The actual value depends on wavelength. For the MM
  fiber, the core diameter is specified instead of the MFD.}
\label{tbl:fibers}
\end{table}

\subsection{Optimizing the dot milling process}
\label{sec:Optimizing}

Our fabrication goal was to obtain spherical mirror profiles with
ROCs in the $1\,$mm range, and with a useful diameter of
$100\,\mu$m. We have achieved good results using identical CO$_2$ beam
parameters for all dots in the pattern. For the results shown below,
the beam had a $1/e^2$ radius $w=140\,\mu$m, and its power was
$P=2.3\,$W. The pulse length $\tau$ was adjusted such that an
individual milling pulse near the center of a flat (cleaved) MM fiber
yielded a depression with depth $t\sim 100\,$nm and diameter
($2\sigma$ of a Gaussian fit) of $\sim 30\,\mu$m. This led to
$17\,\mbox{ms}\le\tau\le25\,\mbox{ms}$ depending on the fiber
type. This dot size is small enough to produce features with the
required resolution and using about one hundred pulses, while a smaller
beam diameter would require more pulses with no obvious
advantage.

\begin{figure}[htbp]
\centering\includegraphics[width=12cm]{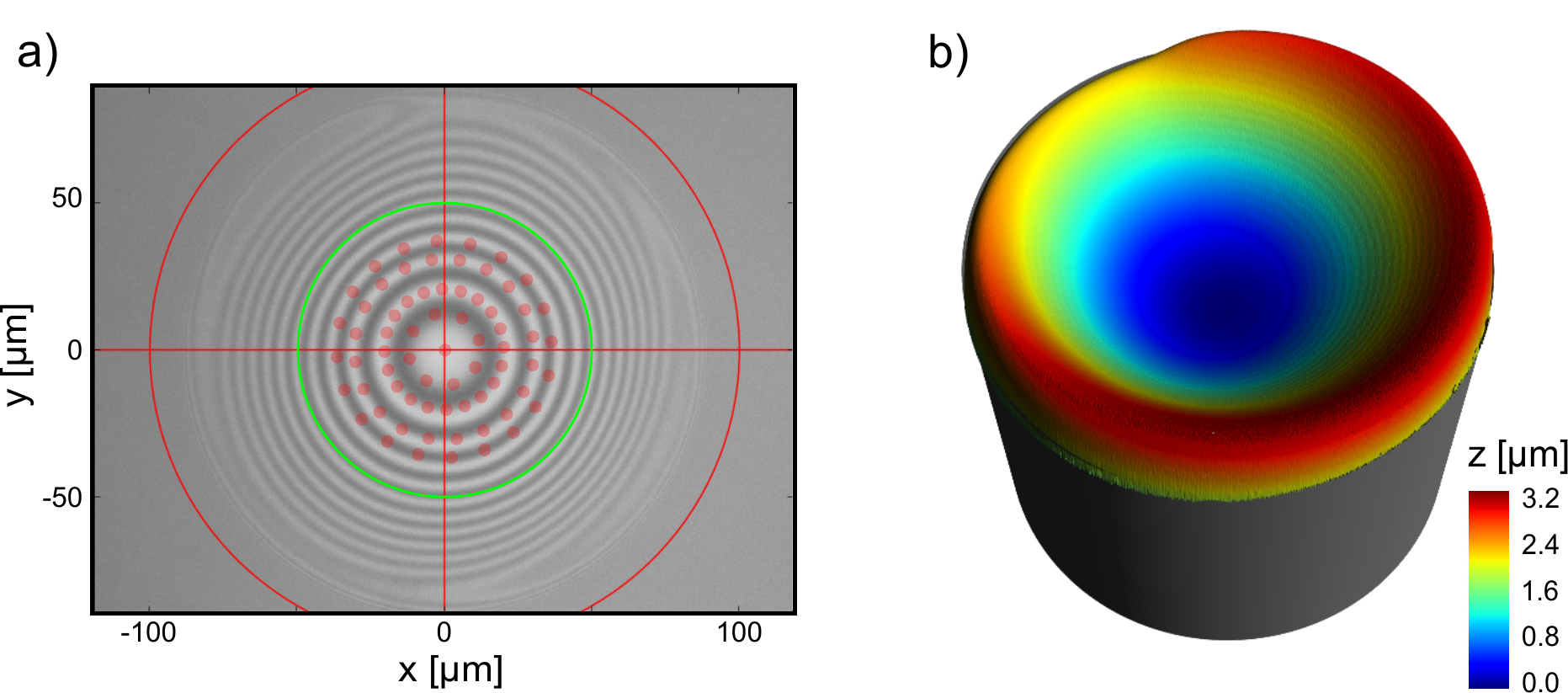}
\caption{Large spherical surface machined on a $200\,\mu$m diameter SM
  fiber.  (a) Interferogram of the surface after processing. The red
  circle shows the initial fiber diameter, the green circle shows the
  area over which the structure was optimized.  The crosshairs
  indicate the center of the fiber. The red dots indicate the
  positions of the CO$_2$ pulses (b) Surface profile measured by
  phase-shifting interferometry. (The grey area was added to indicate
  the fiber orientation.) }
\label{fig:Multishoot}
\end{figure}

The milling process is highly nonlinear, so that the time order of the
pulses also affects the result. Furthermore, due to the finite size of
the fiber, the effect of a given pulse also depends on its distance
from the center of the fiber surface. Therefore, finding a suitable
milling pulse pattern requires some empirical testing to reduce the 
parameter space before a systematic optimization can be done. 
A typical pulse pattern is shown in fig.~\ref{fig:Multishoot}~a); 
our optimization strategy will be discussed in more detail in 
\cite{Garcia15}.
Once the pattern is optimized, processing and characterizing a single 
fiber with a typical 75-dot pattern takes less than 2 minutes.

\begin{figure}[htbp]
  \centering\includegraphics[width=12cm]{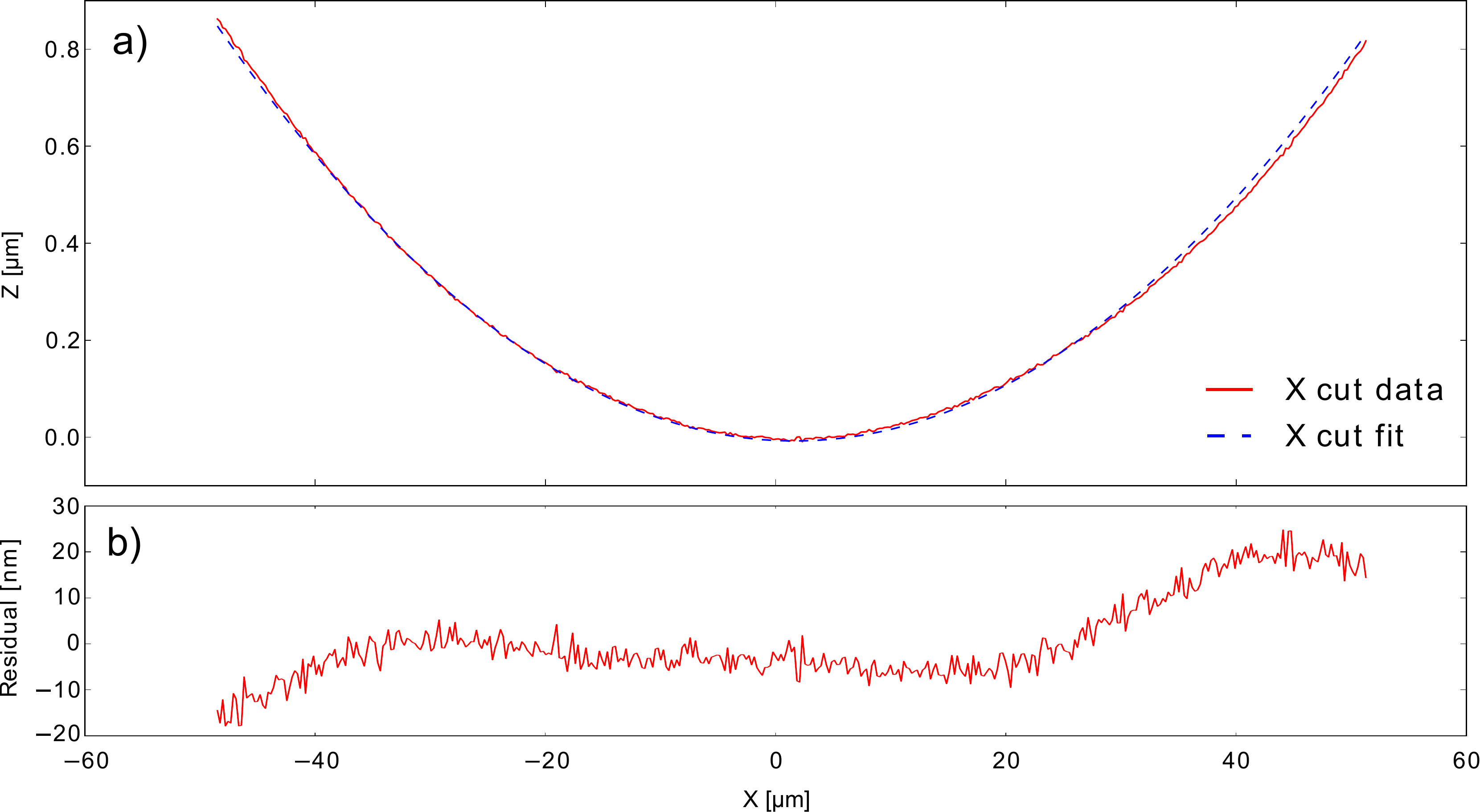}
 \caption{a) Cut along the x-axis of a spherical profile on a
   $220\,\mu$m diameter MM fiber and corresponding cut of a
   two-dimensional spherical fit ($R=1475\,\mu$m) on a circular region
   with a diameter of 100 $\mu$m. b) Residual of the fit. }
 \label{fig:profileCut}
\end{figure}

Using this type of pattern, we have produced mirror profiles on SM, MM
and PC fibers. Figs.~\ref{fig:Multishoot} and \ref{fig:profileCut}
show sample results. The shot positions and parameters were optimized
for $R = 1500\,\mu$m, where the mean deviation from a 2D spherical fit
is smaller than 12\,nm, and the maximum deviation smaller than 40\,nm,
over a circular region of interest with a diameter of
$100\,\mu$m. Between samples processed with identical parameters, the
measured $R$ varies by about $\pm\,10\,\%$ on SM and MM fibers, and by
about $\pm\,20\,\%$ on PC fibers.  Within some limits, larger and
smaller profiles can be machined by simply changing the pulse length
and scaling the shot positions, without optimizing anew.  When
machining smaller $R$ with the profile of Fig.~\ref{fig:Multishoot},
the mean deviation increases up to 59\,nm for $R=330\,\mu m$. When
producing $R > 1500\,\mu $m the mean deviation does not increase
significantly, but cleave imperfections start to compromise the
symmetry of the structure.
		
\begin{figure}[htbp]
  \centering\includegraphics[width=10cm]{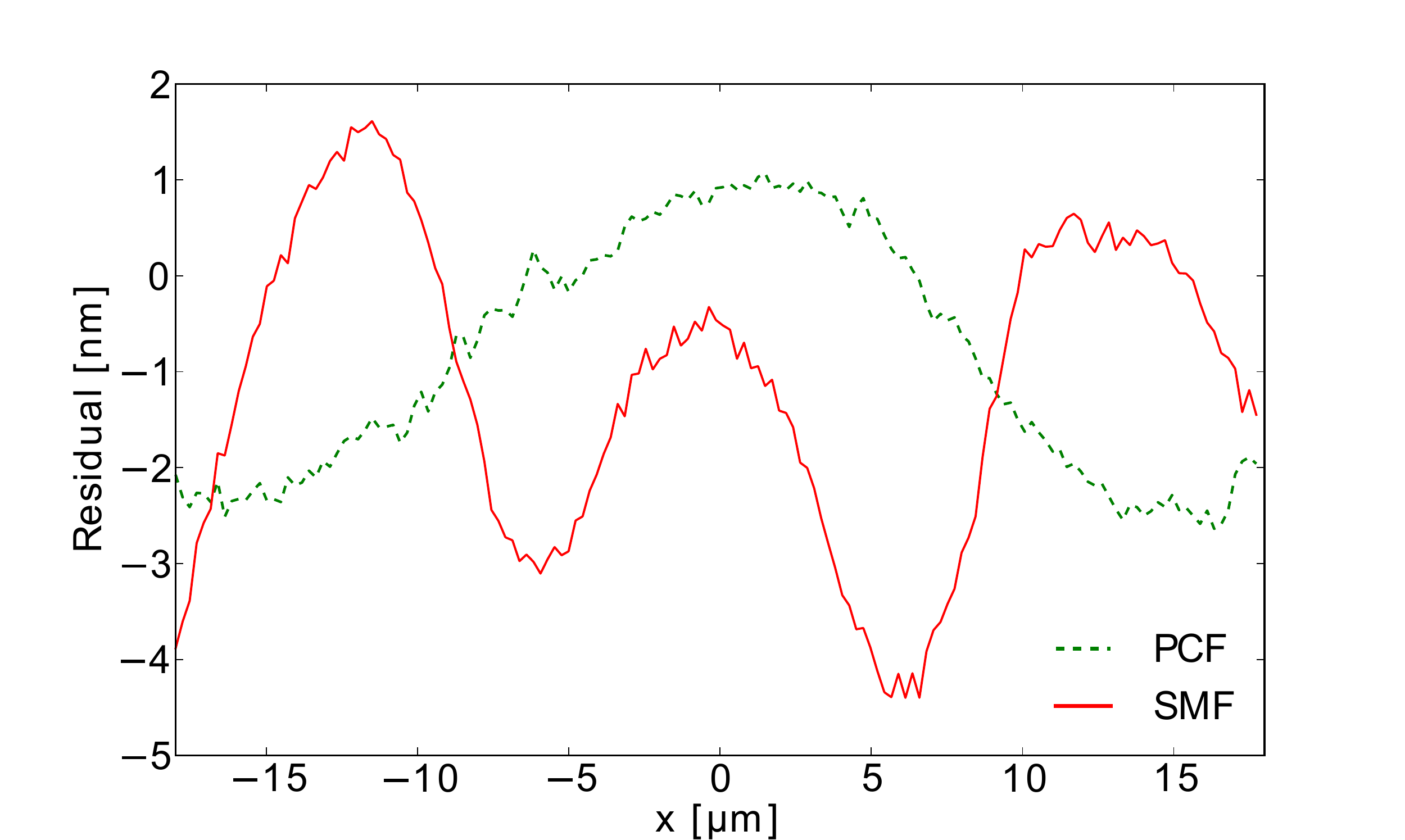}
  \caption{Influence of the doped core region. Shown are the residuals
    of a 2D spherical fit for machined SM and PC fiber surfaces. The
    milling patterns are very similar for both fibers and are close to
    the one shown in {Fig.~\ref{fig:Multishoot}(a)}. 73 pulses were
    used for the SM fiber and 70 for the PC fiber; the pulse length is
    $\tau=17.6\,$ms for the SM fiber and $\tau=20\,$ms for the PC
    fiber.  The data is the average of 120 profile measurements for
    each fiber (see Sec.~\ref{sec:numSim}). The radius of the fit
    region was chosen to be $18\ \mu\mbox{m}$ because this is the mode
    radius on the fiber for a cavity of length $\Lcav=1200\ \mu $m and
    a ROC $R=1650\ \mu$m. The fit for the PC fiber gives $R= 1492\mu$m
    and the residual shows only a slow modulation, which could
    probably be further reduced by fine-tuning the pulse length of the
    last, central pulse. By contrast, the residual of the standard SM
    fiber (fitted $R=1508\mu$m) shows a strong variation at the fiber
    center at the interface between core an cladding material. }
  \label{fig:residuals}
\end{figure}

With SM fibers, we also observe a shape deviation of 20-40nm
height located near the interface between the cladding and doped core,
similar to that reported in \cite{Takahashi14} for several pulses 
on the fiber center.  It can be
compensated to some extent by slight adjustments in the milling
pattern, with the results shown in Fig.~\ref{fig:residuals}. 
Collapsed PC fibers do not
experience this effect, which makes them particularly suitable for
CO$_2$ processing. Likewise, large-core MM fibers are also easy to
process because of their large, homogeneous central region.

\section{Long Fiber-Fabry Perot Cavities}
\label{sec:LongFFPs}

We have machined a large number of SM, MM and PC fibers and had them
coated with an ion-beam sputtered (IBS), dual-wavelength
high-reflectivity coating \cite{Laseroptik}
for 780\,nm and 1560\,nm. (This particular coating was chosen as
required for our application, where Rb atoms are trapped in a far
off-resonant dipole trap inside the cavity.) Its nominal transmission
is $\mathcal T= 30\,$ppm at 780\,nm and 1560\,nm; 
the sum $\mathcal T+\mathcal L$ is $66\pm 2\,$ppm at 780\,nm and
$34\pm 2\,$ppm at 1560\,nm, as determined from finesse
measurements on short FFP cavities.

\subsection{Finesse and transmission measurement}
\label{sec:Finesse}

\begin{figure}[htbp]
\centering\includegraphics[width=13cm]{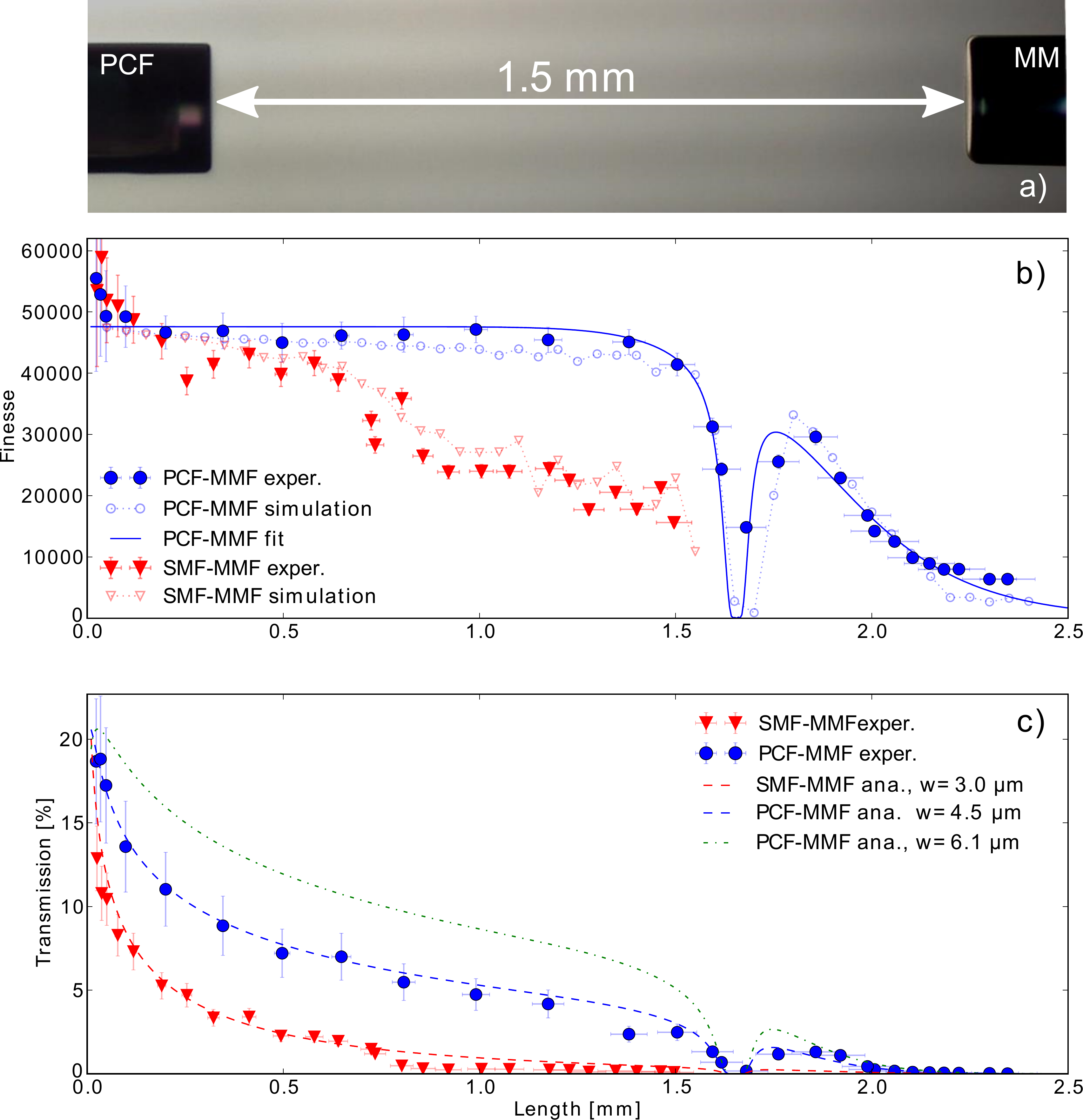}
\caption{(a) Microscope image of a fiber Fabry-Perot cavity with a PC
  fiber (left) and a MM fiber (right). The fibers were illuminated
  from the back to obtain a high contrast for cavity length
  measurement. The collapsed region of the PC fiber is visible. (b)
  Experimental results (filled symbols) and simulations (empty symbols
  with dotted lines, see \ref{sec:numSim}) for the finesse as a
  function of length. Results are shown for an SM-MM cavity (red
  triangles) with $R_1=1508\pm 65\,\mu$m and $R_2=1629\pm 73\,\mu$m
  and for a PC-MM cavity (blue circles) with $R_1=1492\pm 110\,\mu$m
  and $R_2$ as before. (The MM fiber is the same in both cavities.)
  The blue solid line shows the result of the analytical clipping loss
  formula (\ref{eq:lossmodel}) fitted to the PC-MM data. The fit
  parameters are given in Table~\ref{tbl:ROCs}.  (c) Transmission of
  the cavities. The PC fiber improves transmission by more than an
  order of magnitude for $L>1\,$mm. The lines show the calculated
  transmission using eq. \ref{eq:Transmission} and the mode field
  radius $w_f$ given in the legend. 
  $w_f=6.1\,\mu$m  is the mode field radius measured
  independently for the PC fiber (see Sec. \ref{sec:cavTrans}). The
  blue dashed line for $w_f=4.5\,\mu$m fits the PC data well, but the
  large deviation from the nominal value of the PC mode field radius
  remains unexplained. The sharp drop in finesse and transmission
  around $\Lcav=1650\,\mu$m corresponds to the unstable region
  $R_1<\Lcav<R_2$ of the slightly asymmetric cavity.  }

\label{fig:1504umCavityPicture}
\end{figure}

We have built high-finesse cavities from fiber pairs of different
fiber types using a three-axis translation stage equipped with piezo
actuators (Thorlabs MAX311D/M) and a two-axis rotation stage (Thorlabs
PR01/M, GNL10/M).  Several PC-MM cavities were tested at both
wavelengths. Compared to cavities with the standard
Gaussian-shaped fiber mirrors, the difference was striking. With
mirror ROCs on the order of 1\,mm, we could readily achieve stable
cavities with sizeable transmission for lengths $\Lcav>1\,$mm -- a
length never achieved with the Gaussian-shaped mirrors
(Fig.~\ref{fig:1504umCavityPicture}(a)). 

The cavities performed well at both wavelengths, and the measured
finesse values were consistent with the coating specifications. With
traditional SM fibers, it would be impossible to use the same FFP
cavity with wavelengths so far apart. The optical measurements
presented below were performed using an external-cavity diode laser at
$780\,$nm. The cavity length $\Lcav$ was measured with a simple video
microscope.  The free spectral range (FSR) was calculated from $\Lcav$
as $\nu_{\text{FSR}}=c/2\Lcav$ . The cavity length was then scanned
around this position using the piezo. The linewidth $\delta\nu$ of the
TEM$_{00}$ mode was measured using an electro-optic modulator to
generate sidebands for frequency calibration.

These measurements were repeated for different cavity lengths over the
full stability range. Cavity transmission on resonance was also
measured for every length.  In order to eliminate the uncertainty
associated with coupling a free-space beam into the cavity fiber, we
have first injected the free-space beam into an open-ended fiber and
measured the power emerging from the open end before splicing it to
the mirror fiber. In this way, the uncertainty is only given by the
splice. Based on previous splices, we estimate splice transmission to
be $T_{splice}\gtrsim 0.97$ for SM fibers, and $T_{splice}\gtrsim 0.9$
for PC fibers. (The typical value for the PC case is closer to 0.95,
but is very sensitive to all splice preparations, such as the cleave
angle.)

Fig.~\ref{fig:1504umCavityPicture}(b) and (c) show the finesse $\Fi =
\nu_{\text{FSR}}/\delta\nu$ and resonant transmission $T_c$ measured
at 780\,nm as a function of cavity length for two different cavities:
one with an SM fiber (ROC measured by optical profilometry:
$R_1=1508\pm 65\,\mu$m) on the incoupling side and a MM fiber
($R_2=1629\pm73\,\mu$m) for outcoupling; the other with the same MM
fiber, but a PC fiber ($R_1=1492\pm110\,\mu$m) on the incoupling
side. Because the ROCs are not exactly equal, an unstable region
exists for both cavities where $R_1<\Lcav<R_2$. This is clearly visible in
the finesse and transmission data. Comparing the performance of the
two cavities, the PC-MM configuration is superior in every
respect. It shows higher transmission, especially for large $\Lcav$. This
was expected and validates the choice of the PC fiber, although the
experimental values do not reach the theoretical optimum yet (see
Sec.~\ref{sec:cavTrans} below).  Furthermore, with this cavity, $\Fi$ is
almost constant up to $\Lcav\sim 1.5\,$mm, whereas with the SM-MM cavity,
it decreases significantly with $\Lcav$, even for short lengths. Such
a decrease has been observed for all FFP cavities involving SM fibers
\cite{Benedikter15,Takahashi14,Brandstaetter13}; the authors of 
\cite{Takahashi14} have conjectured that it is likely to be caused by 
the ridge in the mirror profile (cf.~Fig.~\ref{fig:residuals}). 
We confirm this quantitatively using a
numerical simulation that takes into account the measured mirror
profile, as described below in Sec.~\ref{sec:numSim}. This adverse
effect is virtually absent with PC fibers, giving them an additional
advantage in FFP cavities.

\subsection{Analytical model: Clipping loss}
\label{sec:analytMod}

Clipping loss for Gaussian cavity modes can be described with a simple
analytical model \cite{Hunger10b}.  The finesse of a high-finesse
Fabry-Perot cavity as a function of the mirror properties is
\begin{equation}
  \mathcal{F} = 
  \frac{2\pi}{\mathcal{L}_A+\mathcal{L}_{S}+\mathcal{L}_C+\mathcal{T}}\,
  \label{eq:finesse}
\end{equation}
where $\mathcal{L}_A$ are the absorption losses in the coatings,
$\mathcal{L}_{S}$ the scattering losses, $\mathcal{T}$ the
transmissions, $\mathcal{L}_{C}$ the clipping losses, and each term is
the sum of the contributions of the two mirrors.  Since all our fiber
mirrors were coated in the same run, and all substrates are
CO$_2$-machined fused silica, we assume that all loss terms are the
same for the two mirrors except for clipping loss. The latter depends
on the effective mirror diameter $D_i$ ($i=1,2$) and the mode field
radius $w_i$ on the mirror:
$\mathcal{L}_C=\mathcal{L}_{\text{cl}}\left({w_{1},D_{1}}\right)+\mathcal{L}_{\text{cl}}\left({w_{2},D_{2}}\right)$,
where \cite{Hunger10b}
\begin{equation}
  \mathcal{L}_{\text{cl}}\left(w_i,D_i\right) = 
     exp \left(-2 (D_i/2)^2/w_i^2\right)\,.
\label{eq:lossmodel}
\end{equation} 
$w_i$ is determined by the cavity geometry, i.e., by the two ROCs
$R_i$ and $\Lcav$. In Fig.~\ref{fig:1504umCavityPicture}(b), the
finesse including the loss model of equation \ref{eq:lossmodel} was
fitted to the data of the PC-MM cavity (solid blue line), leaving
$R_{1,2}$ and $D_{1,2}$ as free parameters. The total losses
$\mathcal{L}_A+\mathcal{L}_{S}+\mathcal{T}=66\,$ppm were measured
independently by short-cavity finesse measurements (see
eq.~\ref{eq:finesse}) and are used as fixed parameters. In a perfectly
symmetric cavity, it would be difficult to attribute clipping loss
values to the individual mirrors, since an overestimation of one value
could be compensated by the other. Here, due to the slight difference
of the ROCs, the beam radius is diverging on the PC fiber mirror for
$L < 1650\,\mu$m and on the MM mirror for $L > 1650\,\mu$m, and
individual $\mathcal{L}_{\text{cl}}$ values are obtained with good
confidence.  (To test the reliability of the fit, we have tried to fix
$D_2$ to a value 5\% larger than the best fit, and refit the data with this
restriction. This lead to an increase of $\chi^2$ by
46\%.) The result (solid blue curve in
Fig.~\ref{fig:1504umCavityPicture}(b)) fits the data well, explaining
the sharp drop of $\Fi$ for {$\Lcav\sim 1.5\,$mm} and the decrease for
$\Lcav\gtrsim 1.8\,$mm: in these regions, the mode radii on the
mirrors $w_{1,2}$ diverge, explaining the decreasing finesse by a rise
of the clipping losses.  The ROC values $R_{1,2}$ and the effective
mirror diameters $D_i$ resulting from the fit are shown in Table
\ref{tbl:ROCs} together with the ROCs obtained from the 3D optical
profilometry of the mirrors.
\begin{table}[h]
\centering
\begin{tabular}{ l|c c| c }
  & \multicolumn{2}{|c|}{clipping loss model} & profilometry \\
  & {D$_m$} [$\mu$m] & R [$\mu$m] & R [$\mu$m] \\
  \hline
  PC (m1)&94& $1645\pm$ 60 & 1492$\pm 110$ \\
  MM (m2)&97& $1665\pm60$ & 1629$\pm 73$ \\
\end{tabular}
\caption{Mirror diameters and ROCs deduced from the finesse data for different length (see Fig.~\ref{fig:1504umCavityPicture}(b)). The last column shows the ROCs of 2D spherical fits to the mirror profiles.}
\label{tbl:ROCs}
\end{table}
All values agree within the error margins, which shows the reliability
of our 3D reconstruction, and confirms that the mirrors are well
described by spheres. The initial goal of creating
mirror structures with $D_m\approx100\,\mu m$ is reached for both fiber
types. Note that this diameter was chosen to meet our requirements. It
was not investigated which maximal diameter the multi-shot method can
create.

For the SM-MM cavity, $\Fi$ drops linearly with increasing $\Lcav$ (red
dots in Fig.~\ref{fig:1504umCavityPicture}(b)). The simple clipping
loss model of eq. \ref{eq:lossmodel} does not fit this data with any
reasonable parameters. This indicates that the loss is not related to
clipping on the mirror edges, but has its origin in irregularities of
the machined structures shown in Fig.~\ref{fig:residuals}. This is
confirmed by the numerical simulations which we will now describe.

\subsection{Full simulation of the cavity mode using reconstructed
  mirror surfaces}
\label{sec:numSim}

To gain further insight into the role of structure imperfections, and
to be able to predict the performance of the machined structures
without building a cavity or even applying a coating, we have
performed numerical simulations of the cavity eigenmodes using the
measured surface profiles.  We have used the FFT toolbox OSCAR
\cite{Degallaix10}. This toolbox simulates cavity eigenmodes for arbitrary
mirror profiles, which are represented as 2D arrays of height
information.  Each of the profiles used in our simulation is the
average of 120 profiles of the same fiber taken in succession to
reduce noise. Apart from this averaging, no fit or filter was used to
process the profile data.  The profiles here were taken after coating,
but no significant deformation with respect to the profiles of the
uncoated structures could be seen.

The alignment procedure for the simulation is analogous to that of a
real cavity: the two profiles are spaced by a fixed distance $\Lcav$, and
the resonance closest to the target cavity length is found. By tilting and translating one of the
profiles, the transmission signal is optimized. In our simulations, we
have not attempted to rotate the profiles around the cavity
axis. Their relative angle is left at an arbitrary value and is not
changed in the optimization. This is justified when the deviation from
rotational symmetry is small, as is the case here. 
The simulations are carried out with higher mirror transmission ($\mathcal T_{sim}=0.01$) compared to the experiment to reduce the computation time. This has no effect on the simulated clipping and scattering losses of the mirrors. The losses and transmission of the coating, which are not contained in the simulation, are experimentally determined by short-cavity finesse measurements and added to the simulated diffraction losses to determine the finesse using Eq. \ref{eq:finesse}.
The results of the simulated finesse are shown in
Fig.~\ref{fig:1504umCavityPicture}(b) as empty symbols connected by dotted lines for both
cavities. Simulation and experiment are in excellent agreement, which
confirms that the linear decay of the finesse for the standard
single-mode 
fiber is explained by the structure itself. Additional loss
effects from coating variations (as discussed in
\cite{Roy11,Brandstaetter13}) are not significant here.  Furthermore,
this remarkable agreement means that such simulations could be used to
optimize structures produced by laser dot milling without laborious
iterations of coating and cavity construction.

\subsection{Cavity transmission}
\label{sec:cavTrans}

To compare the measured cavity transmissions to theoretical
expectations,we have used the simple model described in
\cite{Hunger10b}. The power coupling efficiency $\mmeff$ between the
input fiber and the cavity mode (eq.~\ref{eq:powerTransmissivity})
limits the resonant transmission $T_c$ of a symmetric FFP cavity as
\begin{equation}
  T_c=\mmeff \frac{\mathcal T^2}{(\mathcal T+\mathcal L)^2}\,,
\label{eq:Transmission}
\end{equation}
where $\mathcal L=\mathcal L_S+\mathcal L_A+\mathcal L_C$ is the sum
of all mirror losses.  (We are assuming that the coupling to the MM
output fiber is perfect, which is reasonable because of the large mode
area and high acceptance angle of the MM fiber.)  The lines in
Fig.~{\ref{fig:1504umCavityPicture}(c)} are calculated using this
equation. The $\mathcal T$ values are the same for all three curves
and are those of the multilayer coating (see
Sec.~\ref{sec:analytMod}).  The length-dependent $\mathcal L$ values
are deduced from the finesse measurements and take the additional loss
with increasing $\Lcav$ into account.  The coating properties
determine the maximum $T_c=0.207$, achieved for short cavity
length. The red dashed line uses the nominal mode field radius
$w_f=3\,\mu$m of the SM fiber and fits the SM-MM data well for lengths
up to $\Lcav\approx 770\,\mu$m. The sharp drop at this length is not
expected from Gaussian mode overlap and could be due to imperfections
of the mirror profile discussed above. In that case, optimization of
this profile would bring $T_c$ into agreement with the prediction also
for larger $\Lcav$. However, this would still be more than five times
smaller than the measured $T_c$ of the PC-MM cavity. This confirms the
advantage of the PC fiber. Nevertheless, headroom for further
improvement remains: indeed, the PC-MM result falls below the expected
value if we assume the catalog value for the mode field radius of the
PC fiber ($w_f=8.2\,\mu$m). A good fit is obtained for a much smaller
$w_f=4.5\mu$m (blue dashed line in
Fig.~\ref{fig:1504umCavityPicture}). To
investigate the disagreement more closely, we have measured the mode
field radius of the PC fiber independently. The beam profile was
imaged with a camera at several distances (1.5\, mm - 2.5\,mm) from
the fiber output, to determine the divergence $\theta$ of the beam. By
using $w_f=\lambda/(\pi\theta)$, we found $w_f=(6.1 \pm 0.2)\mu$m,
significantly smaller than the catalog value. The calculated
transmission for this $w_f$ is shown in
Fig. \ref{fig:1504umCavityPicture} (green dash-dotted line), it
remains above the measured values. The fact that the PC fiber mode
is not strictly Gaussian may explain some of the deviation. To
investigate other possible sources, we have
tentatively added tilt and displacement in the calculation of $\mmeff$
following \cite{Joyce84}, but no parameter set could be found which
fits the experimental data as well as the model without displacement
or angle and $w_f=4.5\mu$m. Simple propagation of the outcoupled beam
in the collapsed part (assumed homogenous) changes the coupling to the
cavity mode only slightly and does not explain the difference between
theory and experiment. Further possible effects of wavefront
distortions, diffraction or lensing at the position of the melted
holes of the PC fiber remain to be investigated. Understanding these
effects may lead to further improvements of coupling and transmission.

\section{Conclusion}

These results show that CO$_2$ laser dot milling and large mode area
photonic-crystal fibers form a powerful combination. The maximum
length is extended into the millimeter range while maintaining the
advantages of a miniature, robust, fiber-based approach and an
acceptable overall transmission.  Our results also indicate that an
improved hole-collapsing process is likely to further improve the
transmission. The finesse $\Fi\sim 50000$ reached in our data is
limited by the dual-wavelength coating. The surface roughness of the
CO$_2$ process admits still higher values, and it will be interesting
to see whether these can also be reached at the length scale
introduced here. For a state-of-the-art single-wavelength coating,
total absorption and scattering losses
$\mathcal{L}_A+\mathcal{L}_{S}=2\times13.5\,$ppm have been measured
for a pair of fiber mirrors \cite{Uphoff15}.  Choosing a transmission
of the same value, and adding the clipping losses of 5.5 ppm per
mirror estimated by the simulations shown above for a cavity of
$\Lcav=1\,$mm, it should be possible to produce FFP cavities of
$\Lcav=1\,$mm and a finesse of up to 97,000 with the method presented
here.  Beyond the cavity QED applications for which we have originally
developed it, we note that the free spectral range for a 1.5\,mm fiber
cavity is is 100\,GHz, approaching an interesting range for filtering
applications.

\section{Acknowledgements}
This work was supported by the European Research Council (ERC) (EQUEMI
project, GA 671133) and by the EU Information and Communication
Technologies programme (QIBEC project, GA 284584). K.~O.\ gratefully
acknowledges support from DGA and CNES, and K.~S. from the Army
Research Laboratory Center for Distributed Quantum Information and
FWF. The authors thank Yves Colombe for advice on PC fibers, Tobias
Gross of Laseroptik GmbH for the detailed discussions, Nabil Garroum
and the ENS Physics Department workshops for machining fiber holders,
and Tracy Northup for critical reading of the manuscript.
\end{document}